\documentclass[journal]{IEEEtran}
                                                      \usepackage{blindtext}
\usepackage{graphicx}
\usepackage[utf8]{inputenc}
\usepackage{graphics} % for pdf, bitmapped graphics files
\usepackage{epsfig} % for postscript graphics files
\usepackage{mathptmx} % assumes new font selection scheme installed
\usepackage{amsmath} % assumes amsmath package installed
\usepackage{amssymb}  % assumes amsmath package installed
\usepackage{cite}
\usepackage{physics}
\usepackage{multicol}
\usepackage{mathtools, cuted}
\usepackage[cmintegrals]{newtxmath}
\usepackage{epstopdf}
\usepackage{graphics} % for pdf, bitmapped graphics files
\usepackage{epsfig} 

\usepackage{mathptmx} % assumes new font selection scheme installed
\usepackage{amsmath} % assumes amsmath package installed
\usepackage{amssymb}  % assumes amsmath package installed
\usepackage{cite}
\usepackage{multicol}
\usepackage{mathtools, cuted}
\usepackage[cmintegrals]{newtxmath}
\usepackage{mathtools}

\usepackage{mathptmx}
\usepackage{helvet}
\usepackage{courier}
\usepackage{subfigure}
\usepackage{type1cm}
\usepackage{titlesec}
\usepackage{epsfig} % for postscript graphics files
\usepackage{mathptmx} % assumes new font selection scheme installed
\usepackage{amsmath} % assumes amsmath package installed
\usepackage{amssymb}  % assumes amsmath package installed
\usepackage{cite}
\usepackage{multicol}
\usepackage{mathtools, cuted}
\usepackage[cmintegrals]{newtxmath}
\usepackage{makeidx}         % allows index generation
\usepackage{algorithm}% http://ctan.org/pkg/algorithm
\usepackage{algpseudocode}% http://ctan.org/pkg/algorithmicx
\DeclareMathOperator*{\argmin}{\arg\!\min}

\usepackage{multicol}        % used for the two-column index
\usepackage[bottom]{footmisc}% places footnotes at page bottom
\usepackage{caption}
\usepackage{nomencl}
\usepackage[usenames, dvipsnames]{color}
\usepackage{url}
\usepackage{siunitx}
\usepackage[utf8]{inputenc}

\usepackage{multicol}        % used for the two-column index
\usepackage[bottom]{footmisc}% places footnotes at page bottom
\usepackage{caption}
\usepackage{titlesec}
\usepackage{nomencl}
\usepackage[usenames, dvipsnames]{color}
\usepackage{multirow}
\usepackage{adjustbox}
\usepackage{amsthm} 
\theoremstyle{definition}

\theoremstyle{remark}

\title{\LARGE \bf
Aerial-based Crisis Management Center (ACMC)
}

\author{Hossein Rastgoftar and Salim Hariri
\thanks{Authors are with the Department of Electrical and Computer Engineering, 
        Tucson, AZ 85719, USA
        {\{hrastgoftar, hariri\}@arizona.edu}}%
}

\begin{document}

\maketitle
\thispagestyle{empty}
\pagestyle{empty}

%%%%%%%%%%%%%%%%%%%%%%%%%%%%%%%%%%%%%%%%%%%%%%%%%%%%%%%%%%%%%%%%%%%%%%%%%%%%%%%%
\begin{abstract}
Crisis management (CM) for critical infrastructures, natural disasters such as wildfires and hurricanes, terrorist actions, or civil unrest requires high speed communications and connectivity, and access to high performance computational resources to deliver timely dynamic responses to the crisis being managed by different first responders. CM systems should detect, recognize, and disseminate huge amounts of heterogeneous dynamic events that operate at different speeds and formats. Furthermore, the processing of crisis events and the development of real-time responses are major research challenges when the communications and computational resources needed by CM stakeholders are not available or severely degraded  by the crisis. 
The main goal of the research presented in this paper is to utilize Unmanned Autonomous Systems (UAS) to provide Aerial-based Crisis Management  Center (ACMC)  that will provide the required  communications services and the computational resources that are critically needed by first responders.  
  In our approach to develop an ACMC architecture, we utilize a set of flexible Unmanned Aerial Systems (UAS) that can be dynamically composed to meet the communications and computational requirements of CM tasks. The ACMC services will be modeled as  a deep neural network (DNN) mass transport approach to cover a distributed target in a decentralized manner. This is indeed a new decentralized coverage approach with time-varying communication weights.  Furthermore, our analysis proves  the stability and convergence of the proposed DNN-based mass transport  for a team of  UAS (e.g., quadcopters), where each quadcopter uses a feedback nonlinear control to independently attain the intended coverage trajectory in a decentralized manner.

\end{abstract}

%%%%%%%%%%%%%%%%%%%%%%%%%%%%%%%%%%%%%%%%%%%%%%%%%%%%%%%%%%%%%%%%%%%%%%%%%%%%%%%%
\vspace{-0.5cm}
\section{Introduction}
Over the past two decades, we have seen huge advancements in AI/ML, mobile, cloud, and UAS technologies. However, traditional information Technology  (IT) architectures are not properly equipped utilize aeiral and unmanned autonomous systems to provide an agile communications and computation infrastructures to keep up with the rapidly evolving application demands, especially in managing natural and malicious disasters where IT infrastructures becomes unavailable or severely degraded to be useful for manging CM tasks.  CM applications  share traits such as mobility, dynamic behavior, nonlinear scaling, and sometimes unpredictable growth based on the size and nature of the inputs being processed. The Service Level Objectives (SLO) of these CM applications are dynamic and unpredictable, necessitating continual provisioning and re-provisioning of the communication and computational resources. Satisfying the CM tasks requirements represents major research challenges, especially when the communications and computational resources needed by CM applications are not available or severely degraded  by the crisis.  

\begin{figure}
    \centering
    \includegraphics[width=0.48\textwidth]{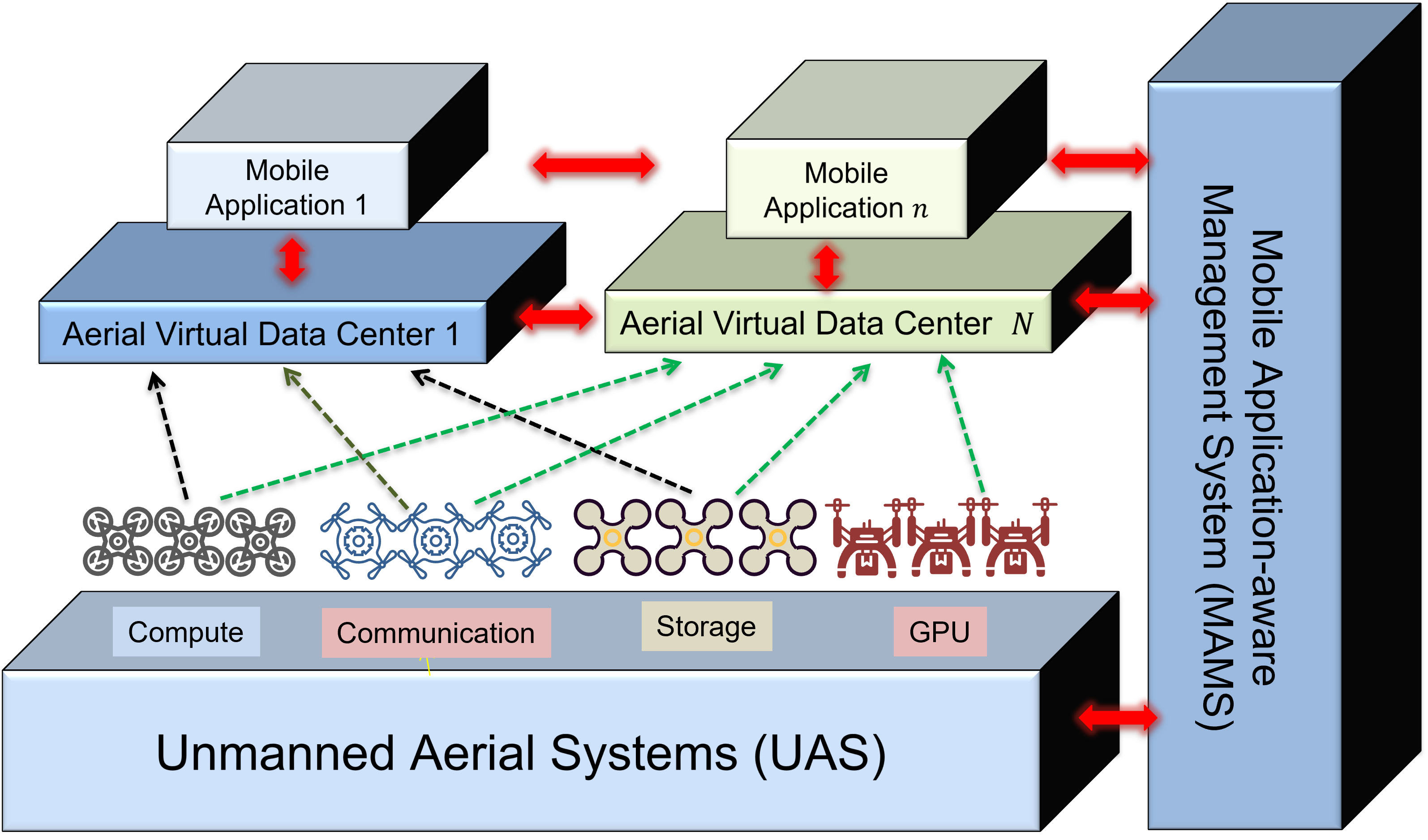}
    \vspace{-0.3cm}
    \caption{The proposed architecture for Aerial-based Crisis Management center (ACMC) for simultaneous provision  computation resources for $n$ applications, defined by $\mathcal{A}=\left\{1,\cdots,n\right\}$, by $N$ Aerial Virtual Data Centers (AVDCs), defined by $\mathcal{V}=\left\{1,\cdots,N\right\}$.}
    \label{schematic}
\end{figure}

This paper aims to develop an innovative Aerial-based Crisis Management Center (ACMC) architecture that utilizes a set of flexible Unmanned Aerial Systems (UAS) that can be dynamically and automatically assembled and re-assembled to meet the dynamic changes in CM applications  (see Fig.  \ref{schematic}). To this end, we will develop aerial-based architectural support, algorithms and modeling techniques to enable the development of aerial composable CM data centers that can efficiently and effectively meet the continuous changes in requirements of mobile data-driven CM applications.

As shown in Fig. \ref{schematic}, we  offer a a mobile application-aware management system (MAMS) that is aware of both the CM application characteristics and the underlying UAS infrastructures to break the barriers between CM applications, middleware/operating system, and UAS hardware layers. The vertical integration of these layers will enable us to build a dynamically customizable Aerial Virtual Data Center (AVDC) that will be optimized to meet the mobile CM application requirements such as QoS, availability, resilience, and energy. Our approach to develop the proposed ACMC architecture is shown in Figure 1 that benefits from virtualization techniques in a variety of ways. Virtual Machines (VMs) have been widely used to provide a layer that is well-positioned in the hardware/software stack of computer systems. This provides fine-grain resource monitoring and control capabilities. In the ACMC approach, we scale the virtualization concept from being at the VM level to an Aerial Virtual Data Center (AVDC) level. For a given mobile application class, the ACMC middleware can build a AVDC that can meet the CM application Service Level Objectives (SLO) such as security, performance, energy consumption, resilience, or availability objectives. The selected AVDC will be then mapped into a set of available UAS resources that will provide the required computations, storage, and communications requirements. The composable UAS building blocks include heterogeneous drones that provide general purpose graphics processing units (GPUs), general purpose processors, storage capabilities and communications services.
\vspace{-0.3cm}
\subsection{Related Work}
ACMC considers UAS as Aerial Virtual Data Centers (AVDCs) that must be optimally and safely configured to provide the best internet coverage over specified areas on the ground. Our proposed method can be considered as a distributed coverage or mass transport problem, which has received a lot of attention by the control community. We indeed propose a novel application for the multi-agent coverage and mass transport problems, while they have previously found  many applications.  Multi-agent coverage has applications in wildfire management \cite{seraj2022multi, khan2022cooperative, haksar2020spatial, gomez2015multi}, border security \cite{pan2022mate}, agriculture \cite{din2022deep, marwah2023analysis}. Also, Mass transport has garnered has found a variety applications in different fields such as brain morphology \cite{gerber2023optimal, ma2019optimal}, image processing \cite{shakib2020mass, song2023chest}, inverse problems \cite{stuart2020inverse}, and cancer detection \cite{lin20213d, wang2010optimal}.

Vononoi-based multi-agent  coverage \cite{bai2021adaptive, nguyen2016discretized, abbasi2017new, luo2019voronoi, adepegba2016multi} has been widely investigated by the control community  for uncovering a distributed target. In \cite{atincc2020swarm, atincc2014supervised}, Voronoi-based multi-agent coverage has been defined as leader-follower and supervised control problems in   which leaders identify zones and followers uncover the target. In \cite{chen2016optimal}, the authors study  the mass transport of linear systems from an initial configuration to an arbitrary target configuration and model it  as an optimization problem involving energy minimization. Furthermore, mass transport has been formulated using linear quadratic Gaussian (LQG) regulation in \cite{hudoba2021discrete}. 

\vspace{-0.5cm}
\subsection{Contributions}
To build the ACMC,  the mass transport method must be scalable and offer the configuration/reconfiguration capability. This requires that a large number of AVDCs can configurably deform over the service area while safety and coverage convergence is guaranteed. To this end, we propose a novel deep neural network- (DNN-) based coverage algorithm plan safe and fast coverage by AVDCs. More specifically, the project offers the following three novel contributions:

\textbf{Contribution 1:} We developed a novel architecture to implement an aerial-based crisis management center to effectively utilize emerging aerial and mobile autonomous devices and systems (e.g., drones, UAS, etc.). The proposed ACMC will provide on-demand and in real-time the critically needed communications services and computational resources for target areas (wildfires, earthquakes, hurricanes, etc.) that IT infrastructures are not available or severely degraded due to natural or malicious disasters.  Furthermore, many other applications can utilize the ACMC services (tactical environments, unmanned autonomous maritime ships, etc.) that might not have needed IT infrastructures and communications services. 

\textbf{Contribution 2:} We use  a DNN to define the inter-ADC communications where the DNN structure is consistent  the reference configuration of the AVDC team, where the reference configuration of the AVDC team and the DNN structure are updated any time the number of AVDCs changes as it is needed due to limited mission duration of UAS.

\textbf{Contribution 3:} We propose a novel ``forward'' method for training the DNN which is fast and convergent as opposed to the existing methods training the neural networks which do not necessarily ensure convergence guarantees. The proposed forward approach will be used to update and obtain the desired position of the AVDCs in real time any time either the coverage map, or the number of AVDCs are updated.

\textbf{Contribution 4:} We define the  AVDC internet provision  as a decentralized multi-AVDC coordination problem with time-varying weights. By considering every AVDC to be a multi-copter drone, we guarantee inter-AVDC stability of the multi-AVDC coordination where we model every AVDC by a nonlinear dynamics recently developed in  \cite{el2023quadcopter}.

% \subsection{Paper Outline}
This paper is organized as follows: Section \ref{Structuring the DNN for Service Provision} provides a solution for updating the DNN structure based on the AVDC reference configuration. Our proposed forward method for training the DNN is presented in Section \ref{Forward Method for Training the DNN}. We define data service provision problem as a decentralized multi-AVDC coordination with time-varying weights in Section \ref{Decentralized Target Acquisition}. Simulation results are presented in Section \ref{Simulation Results} and following by the concluding remarks in Section \ref{Conclusion}.

% \section{Problem Statement}\label{Problem Statement}
\vspace{-0.3cm}
\section{Problem Statement}\label{Structuring the DNN for Service Provision}
We consider an $N$-AVDC team moving above the ground, where we use set $\mathcal{V}=\left\{1,\cdots,N\right\}$ to identify the AVDCs. {\color{black}The AVDC team is deployed to provide internet coverage for $n$ distinct applications where we use set $\mathcal{A}=\left\{1,\cdots,n\right\}$ defines all desired application. We suppose that application $i\in\mathcal{A}$ should be executed over a $2$-dimensional domain with multiple desirable target zones, each of which is geofenced by a ground-based polygon.

We propose to define provision of the communications services and the computational resources as a decentralized multi-agent coordination problem over a finite horizon of time with time-varying communication weights. This problem is decomposed into the following three main sub-problems:}

\textbf{Problem 1: Structuring the Communication Links:} 
Inter-AVDC communications are defined by graph $\mathcal{G}\left(\mathcal{V},\mathcal{E}\right)$, where $\mathcal{E}\subset\mathcal{V}\times \mathcal{V}$ defines the inter-AVDC communication links, i.e. if $(j,i)\in \mathcal{E}$, then, $i\in \mathcal{V}$ communicates with $j\in \mathcal{V}$. Given graph $\mathcal{G}\left(\mathcal{V},\mathcal{E}\right)$, set
\vspace{-0.25cm}
\begin{equation}
    \mathcal{N}_i=\left\{j\in \mathcal{V}:(j,i)\in \mathcal{E}\right\},\qquad \forall i\in \mathcal{V},
\end{equation}
defines in-neighbors of AVDC $i\in \mathcal{V}$.

In this paper, inter-AVDC communication is structured such that graph $\mathcal{G}$ can be represented by by a DNN with $M+1$ layers where we use set $\mathcal{M}=\left\{0,1,\cdots,M\right\}$ to identify the DNN layers. This is indeed beneficial for fast service provision over desired regions, where we can ensure safe coordination of the UAS from an arbitrary initial formation to a desired target configuration while ensuring coverage convergence.  
% The DNN can be formally defined by graph $\mathcal{G}_{DNN}\left(\mathcal{W},\mathcal{E}_{DNN}\right)$, where $\mathcal{W}$ defines the DNN neurons and $\mathcal{E}_{DNN}\subset \mathcal{W}\times \mathcal{W}$ defines the neuron interconnections. 

\textbf{Problem 2: Specifying Communication Weights:} Given the DNN structure, the initial communication weights are unique and obtained based on agents' initial configuration. The DNN is also used to obtain the desired weights based on the distribution of the applications over the ground. Given initial and final DNN (communication) weights, the transient communication weights are defined such that the AVDC team can safely coordinate from their initial configuration toward the desired target configuration.

\textbf{Problem 3: Decentralized Service Provision:} Assuming every AVDC is a multi-copter UAS, we model AVDC motion by a nonlinear dynamics and define the service provision as a decentralized multi-agent coordination problem with time-varying communication weights.
% Next, we  present an approach for specifying inter-AVDC communication by obtaining $\mathcal{G}\left(\mathcal{V},\mathcal{E}\right)$ based on the initial positions of the AVDC team arbitrarily distributed in the motion space, such that it can be converted to a deep neural network. Then, we propose an approach for converting the inter-AVDC communication graph  $\mathcal{G}\left(\mathcal{V},\mathcal{E}\right)$ to a DNN formally structured by $\mathcal{G}_{DNN}\left(\mathcal{W},\mathcal{E}_{DNN}\right)$. 
\vspace{-0.5cm}
\section{Inter-AVDC Communication Structure}
 To ensure that graph $\mathcal{G}\left(\mathcal{V},\mathcal{E}\right)$, defining fixed inter-AVDC communication links, can be converted to a deep neural network, we first divide 
$\mathcal{V}$ into $M+1$ disjoint subsets that are denoted by $\mathcal{V}_0$, $\mathcal{V}_1$, $\cdots$, $\mathcal{V}_{M}$, i.e. $\mathcal{V}$ can be expressed as 
$
\mathcal{V}=\bigcup_{l\in \mathcal{M}}\mathcal{V}_l
$. We can also express set $\mathcal{V}$ as 
\vspace{-0.25cm}
\begin{equation}
    \mathcal{V}=\bigcup_{l\in \mathcal{M}}\mathcal{W}_l,
\end{equation}
where $\mathcal{W}_l$  is related to $\mathcal{V}_l$ by
\vspace{-0.25cm}
\begin{equation}
\mathcal{W}_l=\begin{cases}
\mathcal{V}_l&l\in \{0,M\}\\
\mathcal{V}_l\cup\mathcal{W}_{l-1}&l\in \mathcal{M}\setminus \{0,M\}\\
\end{cases}
.
\end{equation}
We then define in-neighbors of every AVDC $i\in \mathcal{V}_l$ such that the following condition holds:
\vspace{-0.25cm}
\begin{equation}\label{necessity}
    \bigwedge_{l\in \mathcal{M}\setminus \left\{0\right\}}\bigwedge_{i\in \mathcal{V}_l}\left(\mathcal{N}_i\subset \mathcal{W}_{l-1}\right),
\end{equation}
Given $\mathcal{N}_i$, we define
\vspace{-0.25cm}
\begin{equation}
    \mathcal{I}_{i,l}=\begin{cases}
        \mathcal{N}_i&i\in \mathcal{V}_l\\
        \left\{i\right\}&i\in \mathcal{W}_l\setminus\mathcal{V}_l
    \end{cases}
    ,\qquad \forall l\in \mathcal{M}\setminus\left\{0\right\},
\end{equation}
as the set of neurons connected to $i\in \mathcal{W}_l$, from layer $l-1\in \mathcal{M}$. 
% Then, $\mathcal{E}_l\subset \mathcal{W}_{l-1}\times \mathcal{W}_l$ is defined as follows:
% \begin{equation}
%     \mathcal{E}_l=\left\{j\in \mathcal{I}_{i,l}:i\in \mathcal{W}_l,~j\in \mathcal{I}_{i,l}\right\},\qquad\forall l\in \mathcal{M}\setminus\left\{0\right\}.
% \end{equation}

% Inter-AVDC communications can be represented by a deep neural network, 
% % The inter-AVDC communication is structured based on the reference configuration of the AVDC team in $\mathbb{R}^n$, where we use
% % \begin{equation}
% % \mathbf{a}_i=X_{i,j}^0\hat{\mathbf{e}}_j,\qquad i\in \mathcal{V}
% % \end{equation}
% % to express the reference configuration of AVDC $i\in \mathcal{v}$ by a vector form. Note that $\hat{\mathbf{e}}_1$ through $\hat{\mathbf{e}}_n$ are ortho-normal unit vectors. To be more precise, except the $j$-th entry of $\hat{\mathbf{e}}_j$ that is $1$, the remaining entries of $\hat{\mathbf{e}}_j\in \mathbb{R}^{n}$ is $0$, for $j=1,\cdots,n$.

% % \textbf{Mentee AVDC:} Let $\mathcal{V}'\subset \mathcal{V}$ can be expressed as $\mathcal{V}'=\mathcal{V}'_B\cup\mathcal{V}'_I$, where $\mathcal{V}'_B$ and $\mathcal{V}'_I$ are disjoint subsets of $\mathcal{V}$, and $\mathcal{V}'_B$ defines vertices of a convex polytope enclosing $\mathcal{V}'_I$. Then, 
% %  \begin{equation}\label{menteeAVDC}
% %         c(\mathcal{V}'_B)=\argmin_{j\in \mathcal{V}'_I}\left(\sum_{h\in \mathcal{V}'_B}\|\mathbf{a}_{h}-\mathbf{a}_j\|\right),
% %     \end{equation}
% %     obtains the \textit{mentee AVDC} of $\mathcal{V}'_B$ which is located at the closest distance from the boundary AVDCs $\mathcal{V}'_B$.

\textbf{Set $\mathcal{V}_0$:} Given reference positions of the AVDCs, set $\mathcal{V}$ can be expressed as $\mathcal{V}=\mathcal{V}_B\cup\mathcal{V}_I$, where $\mathcal{V}_B$ and $\mathcal{V}_I$ are disjoint subsets of $\mathcal{V}$, and  $\mathcal{V}_B$ and $\mathcal{V}_I$ define ``boundary'' and ``interior'' AVDCs of $\mathcal{V}$, respectively. \textit{In this work, we assume that the AVDC team are contained by a convex  polytope whose vertices are occupied by the boundary AVDCs. }

Given the reference configuration of the AVDC team, we also determine the core leader AVDC. Core leader is an interior AVDC that is at the closest distance from the boundary AVDCs, and assigned by solving the following optimization problem:
\vspace{-0.25cm}
\begin{equation}\label{coreAVDC}
        c(\mathcal{V}_B)=\argmin_{j\in \mathcal{V}_I}\left(\sum_{h\in \mathcal{V}_B}\|\mathbf{a}_{h}-\mathbf{a}_j\|\right).
    \end{equation}
where $\mathbf{a}_i$ is the reference (initial) position of agent $i\in \mathcal{V}$. Set 
\vspace{-0.25cm}
\begin{equation}
    \mathcal{V}_0=\mathcal{V}_B\bigcup c(\mathcal{V}_B)
\end{equation} 
includes the boundary and core AVDCs. Note that $\mathcal{V}_0$'s AVDCs are called leaders because they do not communicate with any other AVDCs to evolve in the motion space. This property of $\mathcal{V}_0$ can be formally specified by
\vspace{-0.25cm}
\begin{equation}
    \bigwedge_{i\in \mathcal{V}_0}\left(\mathcal{N}_i=\emptyset\right).
\end{equation}

\textbf{Sets $\mathcal{V}_1$ through $\mathcal{V}_{M-1}$:} AVDCs defined by $\mathcal{V}_l$, for every $l\in \mathcal{M}\setminus \left\{0,M\right\}$ can act as both followers and in-neighbors. More specifically, $\mathcal{V}_l$'s AVDCs act as followers because their in-neighbors are subsets of $\mathcal{V}_{l-1}$, per the feasibility requirement given by Eq. \eqref{necessity}. On the other hand, every AVDC $i\in \mathcal{V}_l$ is an in-neighbor for at least one AVDC belonging to $\mathcal{V}_{l+1}$, when $l\in \mathcal{M}\setminus \left\{0,M\right\}$.

\textbf{Set $\mathcal{V}_M$:} The AVDC defined by $\mathcal{V}_M$ are pure followers which mean that $\mathcal{V}_M$'s AVDCs update their own positions based on positions of ther in-neighbors, belonging to $\mathcal{V}_{M-1}$, but, they do not act as in-neighbors for any other AVDCs.

 We desire that the reference communication weights are unique and consistent with the AVDC's positions. To this end, we restrict inter-AVDC communication to satisfy the following two requirements:

\textbf{Requirement 1:} Every AVDC $i\in \mathcal{V}\setminus \mathcal{V}_0$ solely communicates with three AVDCs. Therefore,
\vspace{-0.25cm}
\begin{equation}\label{necessity}
    \bigwedge_{l\in \mathcal{M}\setminus \left\{0\right\}}\bigwedge_{i\in \mathcal{V}_l}\left(\left|\mathcal{N}_i\right|=3\right),
\end{equation}
To be more precise, in-neighbors of  $i\in \mathcal{V}\setminus \mathcal{V}_0$ are placed at vertices of a triangle that encloses  $i\in \mathcal{V}\setminus \mathcal{V}_0$. This simplex, is called the \textit{communication simplex} of AVDC $i\in \mathcal{V}\setminus \mathcal{V}_0$.

\textbf{Requirement 2:} If this requirement is satisfied, reference communication weight $\omega_{i,j}$, between AVDC $i\in \mathcal{V}\setminus \mathcal{V}_0$ and $j\in \mathcal{N}_i$, is unique, positive, consistent with reference configuration, and  obtained by solving the following equations:
\vspace{-0.25cm}
\begin{subequations}
    \begin{equation}\label{weightcond1}
    \mathbf{a}_i=\sum_{j\in \mathcal{N}_i}\omega_{i,j}\mathbf{a}_j,\qquad \forall i\in \mathcal{V},
    \end{equation}
       \begin{equation}\label{weightcond2}
    \sum_{j\in \mathcal{N}_i}\omega_{i,j}=1,\qquad \forall i\in \mathcal{V},
    \end{equation}
\end{subequations}

\vspace{-0.3cm}
\section{Forward Method for Training the DNN}\label{Forward Method for Training the DNN}
To train the DNN,  we first generate a heat map to specify the target distribution by   a continuous and differentiable field over a two-dimensional motion space. By knowing the heat map and the DNN structure, we develop a mass-centric approach to update the desired final  position of every AVDC. Lastly, we formulate DNN weights based AVDCs' desired positions in Section  \ref{Communication Weights}.
\vspace{-0.3cm}
\subsection{Service Distribution Heat Map}\label{Target Distribution Heat Map}
% We assume that AVDCs are deployed to provide services for $n$ applications defined . 
For service application $j\in \mathcal{A}$, we let set $\mathcal{D}_j$ define discrete target points where $\mathbf{d}_{i,j}$ denotes the position of target $i\in \mathcal{D}_j$, for $j\in \mathcal{A}$. Then, we define heat map $\mathcal{H}_j:\mathbb{R}^2\rightarrow\mathbb{R}_+$ for $j\in \mathcal{A}$  by the following multi-variate Gaussian distribution:
\vspace{-0.25cm}
\begin{equation}
\resizebox{0.99\hsize}{!}{%
$
    \mathcal{H}_j\left(\mathbf{r},\mathcal{D}_j\right)={1\over \left|\mathcal{D}_j\right|}\left(2\pi\right)^{-1}\sum_{i\in \mathcal{D}_j}\mathrm{det}\left(\mathbf{\Sigma}_i\right)^{-1\over2}\mathrm{exp}\left(-\left(\mathbf{r}-\mathbf{d}_{i,j}\right)^T\mathbf{\Sigma}_i^{-1}\left(\mathbf{r}-\mathbf{d}_{i,j}\right)^T\right),
$
}
\end{equation}
% for every $i\in \mathcal{D}_j$ and $j\in \mathcal{A}$, 
% to specify the likelihood of existence of importance of application $j\in \mathcal{A}$ at $\mathbf{r}\in \mathbb{R}^2$. Note
where $\mathbf{\Sigma}_i$ is the covariance matrix corresponding to target $i\in \mathcal{D}_j$ and  $\alpha_j \in \left[0,1\right]$ is a positive weight to quantify the priority of application $j\in \mathcal{A}$, where 
\vspace{-0.25cm}
\begin{equation}
    \sum_{j\in \mathcal{A}}\alpha_j=1.
\end{equation}
The service distribution heat map (SDHM) is defined by
\vspace{-0.25cm}
\begin{equation}
    \mathcal{H}\left(\mathbf{r}\right)=\sum_{j\in \mathcal{A}}\alpha_j\mathcal{H}_j\left(\mathbf{r},\mathcal{D}_j\right).
\end{equation}

\vspace{-0.3cm}
\subsection{AVDCs' Desired Positions}\label{AVDCs' Desired Positions}
Given the target set distribution, we develop a mass-centric approach to determine and update the desired position of every AVDC $i\in \mathcal{V}$ so that the best coverage of the SDHM is achieved. Desired position of the core AVDC is assigned based on the desired positions of the boundary AVDCs by
\vspace{-0.25cm}
    \begin{equation}
    \mathbf{p}_{c\left(\mathcal{V}_B\right)}=\dfrac{\sum_{i\in \mathcal{V}_B}\mathbf{p}_i}{\left|\mathcal{V}_B\right|}.
\end{equation}
% is considered as the desired position of the core AVDC. 
Therefore, positions of $\mathcal{V}_0$'s AVDCs are known and constant at any time $t$.    % \end{equation}
% \end{assumption}
% The objective of DNN training is that the AVDCs' are distributed such that the best coverage of the dataset $\mathcal{D}(t)=\left\{1,\cdots,n_d(t)\right\}$ is achived.
% The objective of the DNN training is that the final position  $\mathbf{p}_i$ of every AVDC $i\in \mathcal{V}\setminus \mathcal{V}_0$ is defined such that the best coverage of the target is achieved. Assuming the boundary AVDCs are at vertices of convex polytope $\mathcal{P}\subset \mathbb{R}^n$, the we position the core AVDC at $\mathbf{p}_{core}$ the centroid of the target distributed over $\mathcal{P}$. Therefore, 
% \begin{equation}
%     \mathbf{p}_{core}=\dfrac{\int_{\mathcal{P}}\mathcal{H}\left(r\right)d\mathbf{r}}{\int_{\mathcal{P}}d\mathbf{r}}.
% \end{equation}
% Therefore, position $\mathbf{p}_i$ of every AVDC $i\in \mathcal{V}_0$ is known. 
By defining \vspace{-0.25cm}
\begin{equation}\label{Cind}
        \mathcal{C}_i=\left\{\sum_{j\in \mathcal{N}_i}\gamma_j\mathbf{p}_j\subset \mathbb{R}^n:\sum_{j\in \mathcal{N}_i}\gamma_j=1,~\gamma_j\geq 0,\forall j\in \mathcal{N}_i\right\}
    \end{equation}
as the convex hull specified by the in-neighbors of AVDC $i\in \mathcal{V}\setminus \mathcal{V}_0$, desired position of every multi-copter $i\in \mathcal{V}\setminus \mathcal{V}_0$ is obtained by
\vspace{-0.25cm}
\begin{equation}
    \mathbf{p}_{i}=\dfrac{\int_{\mathcal{C}_{i}}\mathcal{H}\left(\mathbf{r}\right)d\mathbf{r}}{\int_{\mathcal{C}_{i}}d\mathbf{r}},\qquad i\in \mathcal{V}\setminus \mathcal{V}_0.
\end{equation}
\vspace{-0.3cm}
\subsection{Communication Weights}
\label{Communication Weights}
% \subsection{Desired Communication Weights}
In-neighbors of every multi-copter $i\in \mathcal{V}$, defined by $\mathcal{N}_i$, are at vertices of a triangle enclosing $i\in \mathcal{V}$. 
% To be more precise, in-neighbors of $i\in \mathcal{V}$ are at vertices of  communication triangle and communication tetrahedron for $n=2$ and $n=3$, respectively, where communication triangle or tetrahedron of AVDC $i\in \mathcal{V}$ encloses $i\in \mathcal{V}$. 
Because of this, we can express (final) desired position $\mathbf{p}_i$ as convex combination of desired positions of its in-neighbors as follows:
% \vspace{-0.25cm}
% \begin{subequations}\label{desiredrawwwwwwwwwwwwwwww}
%     \begin{equation}
%         \mathbf{a}_i=\sum_{j\in \mathcal{N}_i}\omega_{i,j}\mathbf{a}_{j},\qquad \forall i\in \mathcal{V},
%     \end{equation}
%     \begin{equation}
%         \sum_{j\in \mathcal{N}_i}\omega_{i,j}=1,\qquad \forall i\in \mathcal{V},
%     \end{equation}
% \end{subequations}
\vspace{-0.25cm}
\begin{subequations}\label{desiredrawwwwwwwwwwwwwwww}
    \begin{equation}
        \mathbf{p}_i=\sum_{j\in \mathcal{N}_i}\varpi_{i,j}\mathbf{p}_{j},\qquad \forall i\in \mathcal{V},
    \end{equation}
    % \end{subequations}
\vspace{-0.25cm}
    \begin{equation}
        \sum_{j\in \mathcal{N}_i}\varpi_{i,j}=1,\qquad \forall i\in \mathcal{V},
    \end{equation}
\end{subequations}
where  $\varpi_{i,j}$ is the final communication weight  of $i\in \mathcal{V}$ with in-neighbor $j\in \mathcal{N}_i$. The communication weight of AVDC $i\in \mathcal{V}$ with $j\in \mathcal{N}_i$ is then denoted by  $w_{i,j}(t)$ and defined as follows:
\vspace{-0.25cm}
\begin{equation}\label{communicationweighttimevarying}
% \resizebox{0.99\hsize}{!}{%
% $
w_{i,j}(t)=
\begin{cases}
\left(1-\beta(t)\right)\omega_{i,j}+\beta(t)\varpi_{i,j}& t\in \left[0,t_f\right]\\
\varpi_{i,j}&t\geq t_f
\end{cases}
,
% $
% }
\end{equation}
where $t_0$ and $t_f$ denote the reference and final times, respectively; $\omega_{i,j}$ is the initial communication weights obtained by Eqs. \eqref{weightcond1} and  \eqref{weightcond2}; and $\beta:\left[0, t_f\right]\rightarrow \left[0,1\right]$ is an increasing function satisfying   $\beta(t_0)=0$ and   $\beta(t_f)=1$. 
\vspace{-0.4cm}
\section{Decentralized Internet Coverage}\label{Decentralized Target Acquisition}
The internet coverage problem is defined as decentralized multi-agent coordination structured by the DNN,  where the DNN's neurons are operated by differential equations. 
Particularly, neuron $i\in \mathcal{W}_l$ ($l\in \mathcal{M}$) represents AVDC $i\in \mathcal{V}$ where the dynamics of AVDC $i\in \mathcal{V}$ iven by
\vspace{-0.25cm}
\begin{equation}\label{nonlineardynamics}
\begin{cases}
    \dot{\mathbf{x}}_i=\mathbf{F}_i\left(\mathbf{x}_i,\mathbf{u}_i\right)\\
    \mathbf{r}_i=\mathbf{h}_i\left(\mathbf{x}_i\right)
\end{cases}    
\end{equation}
operates neuron $i\in \mathcal{V}$. Note that $\mathbf{x}_i$ and $\mathbf{u}_i$ are the state and input vectors of dynamics \eqref{nonlineardynamics}, respectively, and $\mathbf{r}_i$ denoting position of AVDC $i\in \mathcal{V}$, is considered as the output vector of dynamics \eqref{nonlineardynamics}. The input vector of agent $i\in \mathcal{V}$ is defined by 
\begin{equation}\label{ridd}
    \mathbf{r}_{i,d}=\begin{cases}
    \mathbf{p}_i&i\in \mathcal{V}_0\\
        w_{i,j}\mathbf{r}_j&i\in \mathcal{V}\setminus \mathcal{V}_0\\
    \end{cases}
\end{equation}

\begin{figure}
    \centering
    \includegraphics[width=0.48\textwidth]{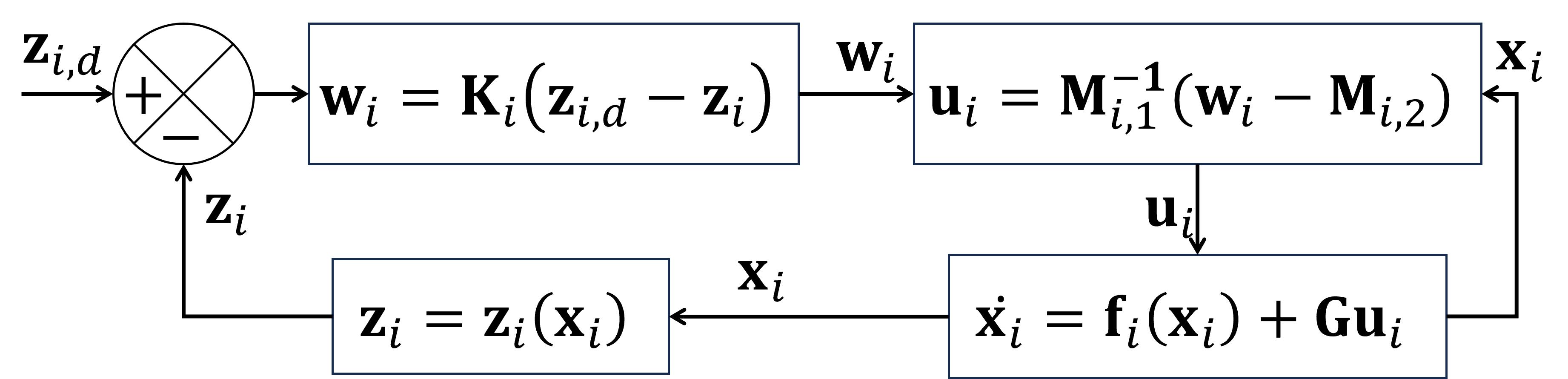}
    \caption{The block diaram of the trajectory tracking control used by every AVDC $i\in \mathcal{V}$.}
    \label{Blockdiagram}
\end{figure}
\begin{figure}
    \centering
    \includegraphics[width=0.48\textwidth]{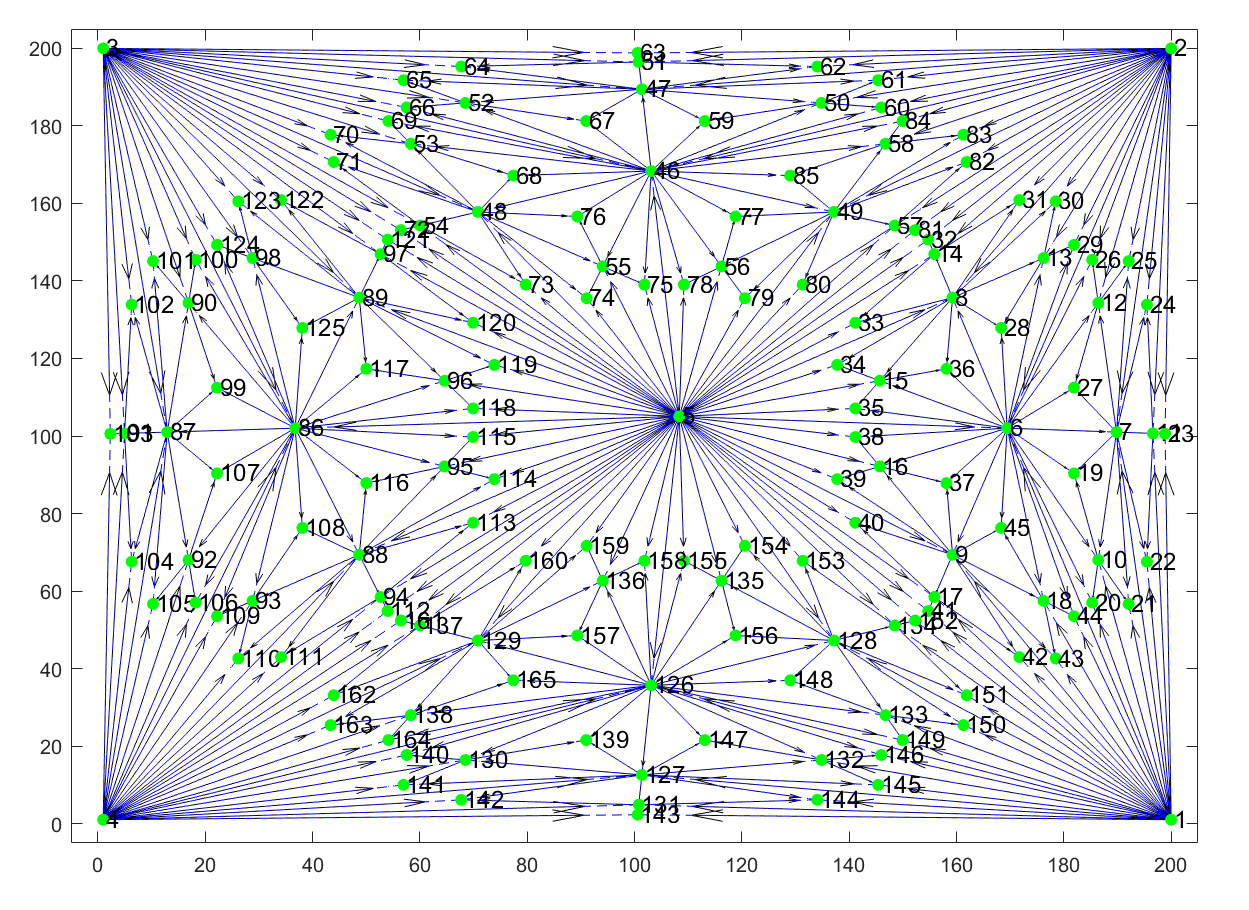}
    \vspace{-0.3cm}
    \caption{The initial formation of the AVDC team and inter-agent communication links.}
    \label{InitialFormation}
\end{figure}
\begin{figure*}
    \centering
    \includegraphics[width=0.98\textwidth]{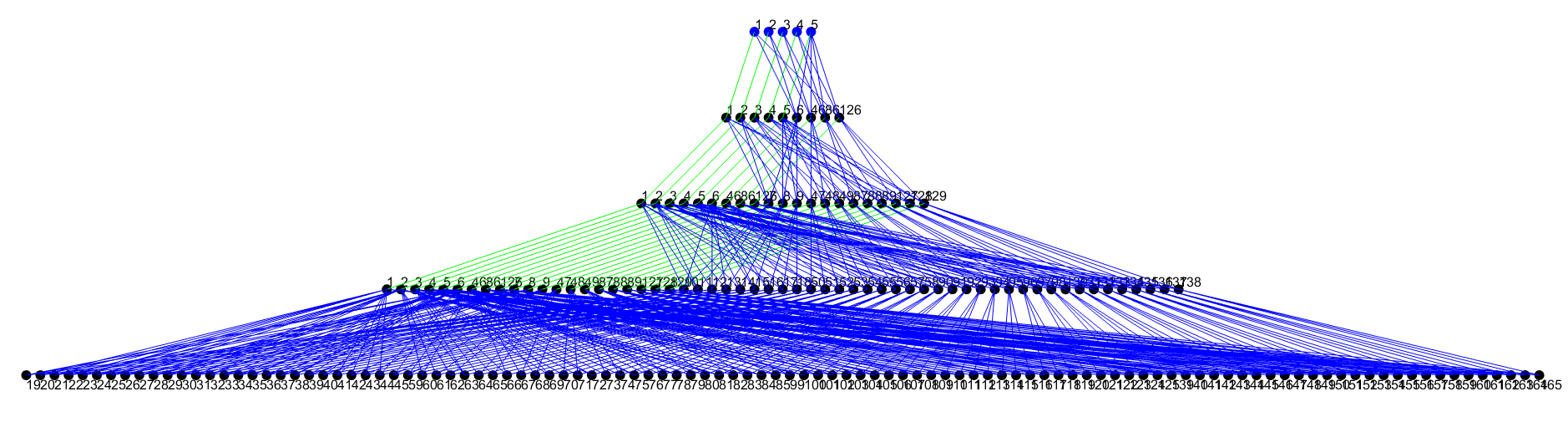}
    \vspace{-0.3cm}
    \caption{The structure of the DNN assigned based on the reference configuration of the AVDC team.}
    \label{DNNStructure}
\end{figure*}

% \begin{figure}
%     \centering
%     \includegraphics[width=0.48\textwidth]{FinalConfiguration.png}
%     \caption{The desired final configuration of the AVDC team, obtained by using the approach presented in Section \ref{AVDCs' Desired Positions},  for covering the desired target contained by the rectangular and triangular domains.}
%     \label{DNNStructure}
% \end{figure}
\begin{figure}
    \centering
    \includegraphics[width=0.48\textwidth]{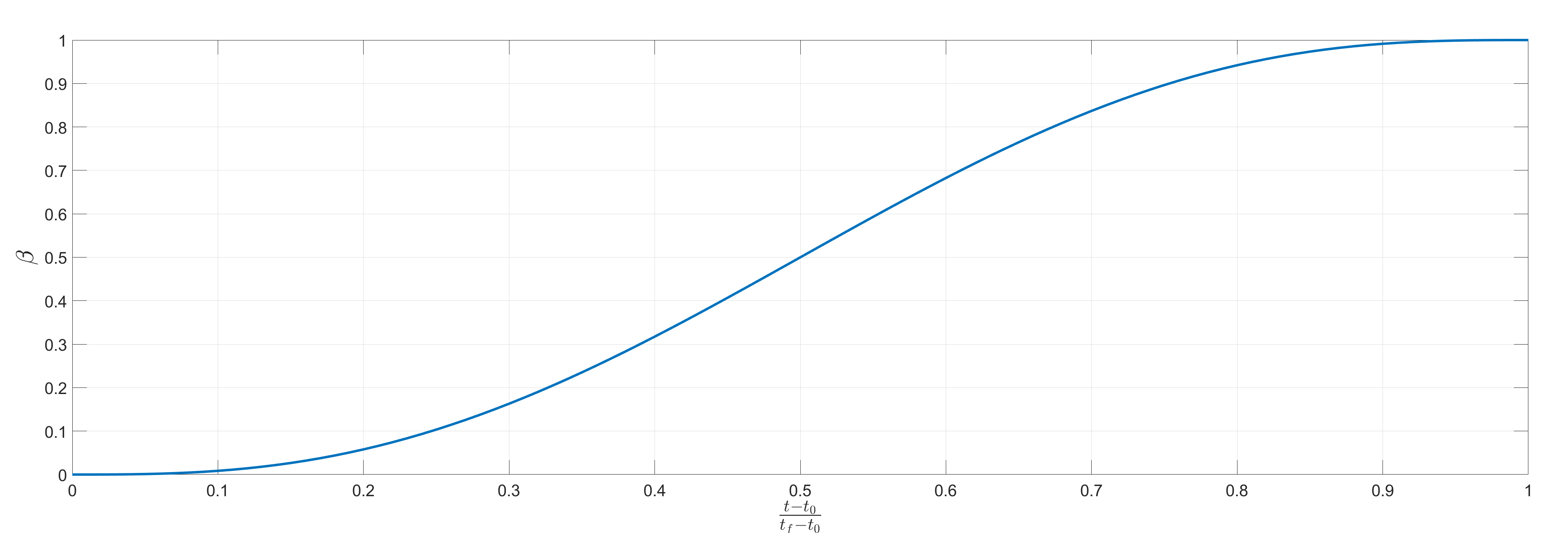}
    \vspace{-0.3cm}
    \caption{The quintic polynomial used for defining $\beta$ versus ${t-t_0\over t_f-t_0}$.}
    \label{beta}
\end{figure}
\begin{figure*}[h]
\centering
\subfigure[$t=15s$]{\includegraphics[width=0.32\linewidth]{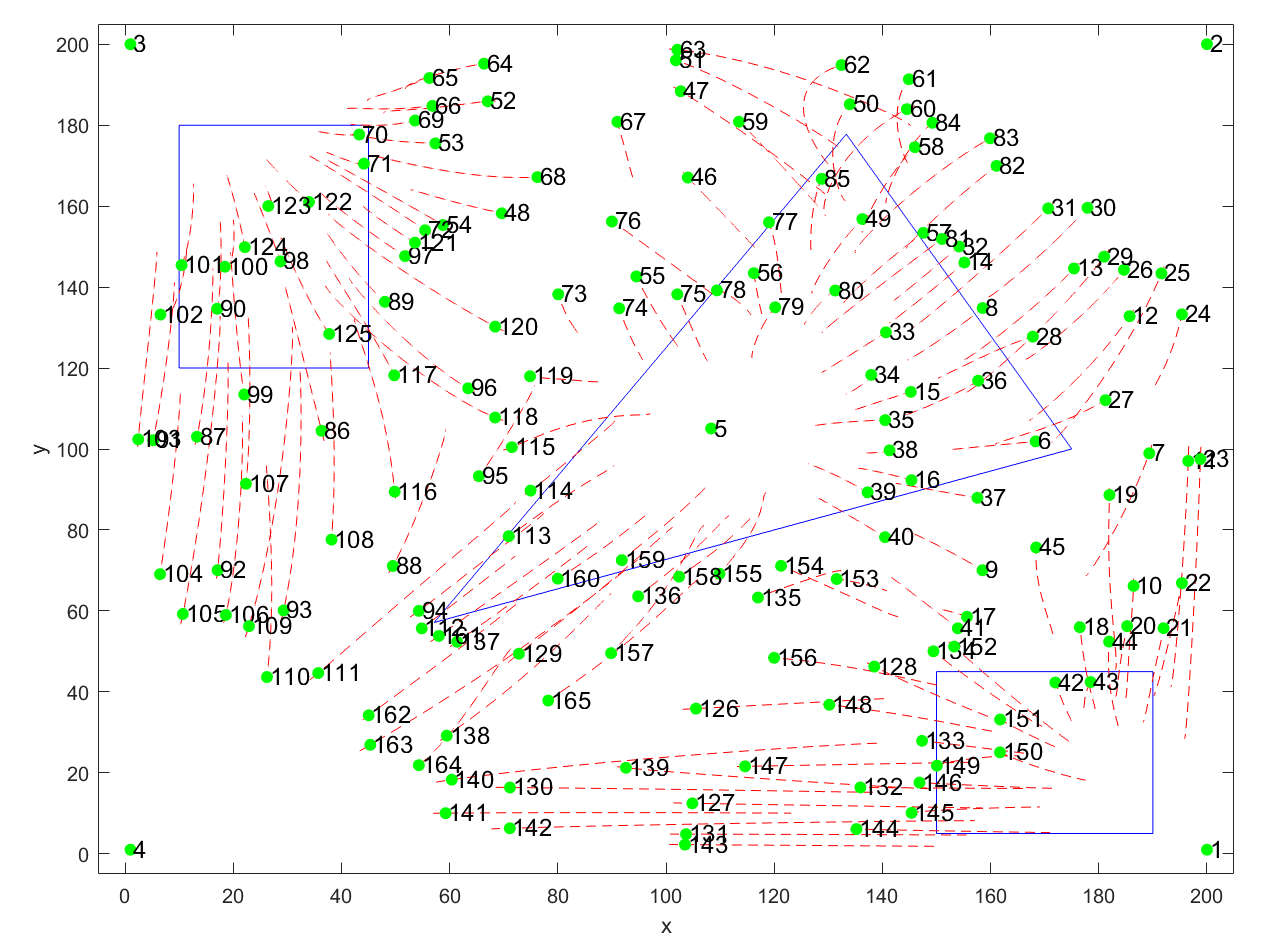}}
\subfigure[$t=35s$]{\includegraphics[width=0.32\linewidth]{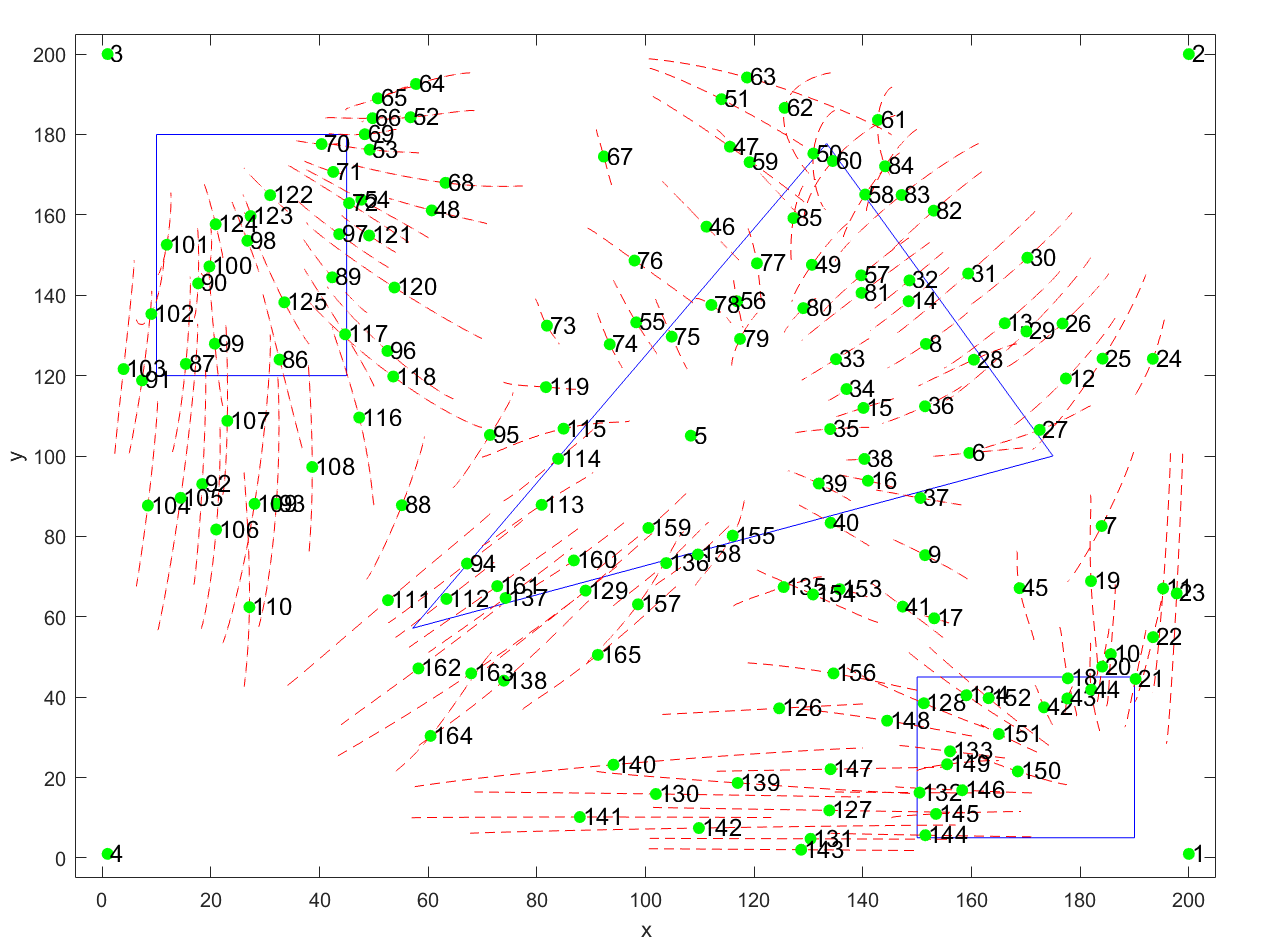}}
\subfigure[$t=80s$]{\includegraphics[width=0.32\linewidth]{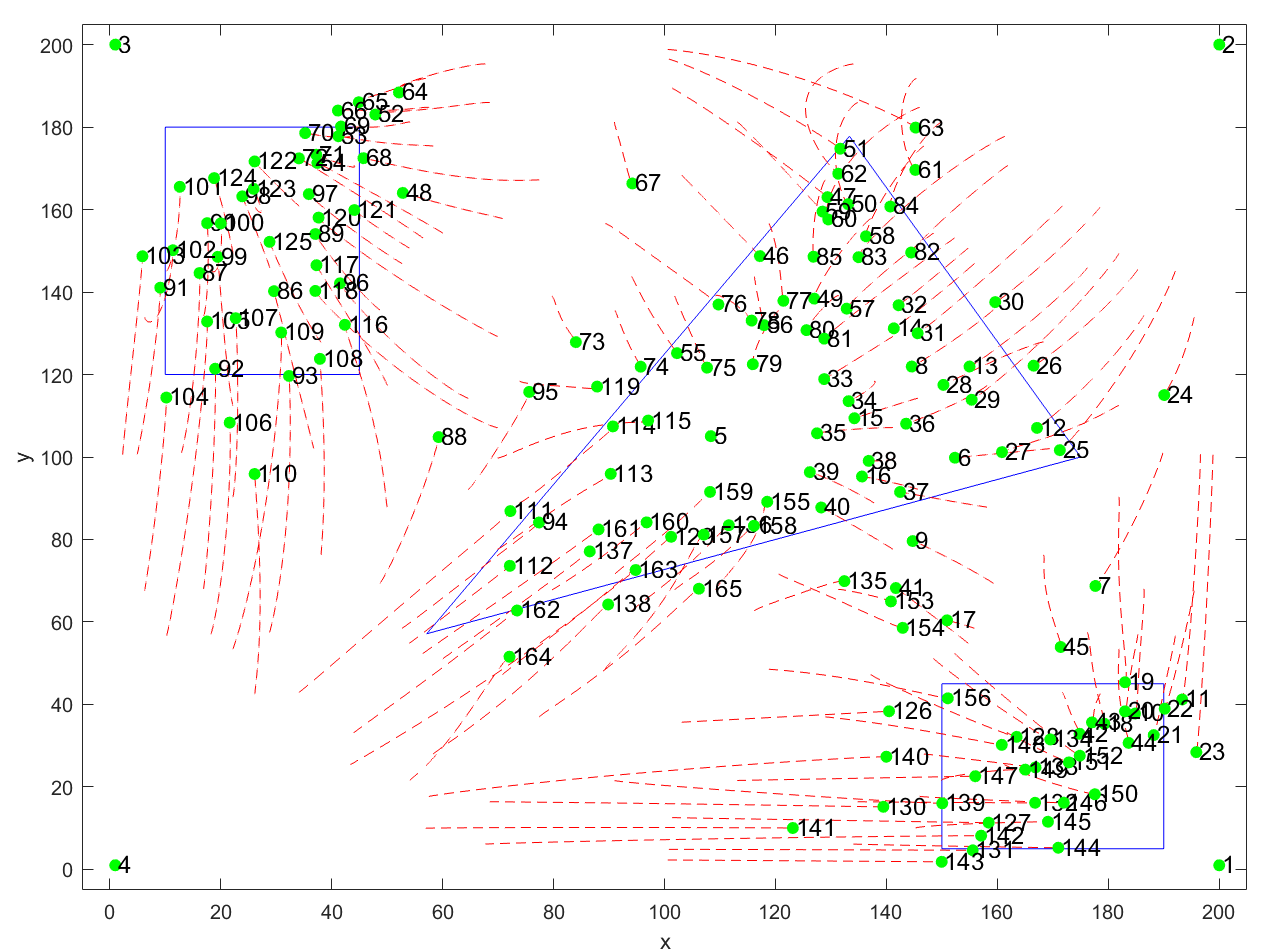}}
% \subfigure[]{\includegraphics[width=0.32\linewidth]{w-37-4-V2.png}}
% \subfigure[]{\includegraphics[width=0.32\linewidth]{w-37-28-V2.png}}
% \subfigure[]{\includegraphics[width=0.32\linewidth]{x37-V2.png}}
% \subfigure[]{\includegraphics[width=0.32\linewidth]{y37-V2.png}}
% \vspace{-0.5cm}
\vspace{-0.3cm}
\caption{An AVDC team consisting of $N=165$ ($\mathcal{V}=\left\{1,\cdots,165\right\}$) UAS aims to provide services for $n=3$ applications ($\mathcal{A}=\left\{1,2,3\right\}$), where each operation is independently occurring in one of the illustrated zones. Intermediate configurations of the AVDC team at (a) $t=15s$, (b) $t=35s$, and (c) $t=80s$.
}
\label{IntemediateConfs}
\end{figure*}
\vspace{-0.5cm}
\subsection{AVDC Motion Model}
Assuming AVDC $i\in \mathcal{V}$ is a multi-copter UAS with mass $m_i$, we use the nonlinear dynamics presented in \cite{rastgoftar2021safe} to obtain $\mathbf{F}_i\left(\mathbf{x}_i,\mathbf{u}_i\right)$. The state vector $\mathbf{x}_i\in \mathbb{R}^{14\times 1}$   is given by 
\vspace{-0.25cm}
\begin{equation}
    \mathbf{x}_i=\left[x_i~y_i~z_i~\dot{x}_i~\dot{y}_i~\dot{z}_i~\phi_i~\theta_i~\psi_i~\dot{\phi}_i~\dot{\theta}_i~\dot{\psi}_i~f_i~\dot{f}_i\right]^T
\end{equation}
where  $x_i$, $y_i$, and $z_i$ define UAS position components at (continuous) time $t$; $\phi_i$, $\theta_i$, and $\psi_i$ define roll, pitch, and yaw angles of AVDC $i\in \mathcal{V}$; and $f_i$ is the magnitude of the  thrust force generated by AVDC $i\in \mathcal{V}$. We also define 
\vspace{-0.25cm}
\begin{equation}
    \mathbf{u}_i=\left[u_{1,i}~u_{2,i}~u_{3,i}~u_{4,i}\right]^T
\end{equation}
where $u_{1,i}=\ddot{f}_i$, $u_{2,i}=\ddot{\phi}_i$, $u_{3,i}=\ddot{\theta}_i$, and $u_{4,i}=\ddot{\psi}_i$. We can express $\mathbf{F}_i\left(\mathbf{x}_i,\mathbf{u}_i\right)$ as
\vspace{-0.25cm}
\begin{equation}
\mathbf{F}_i\left(\mathbf{x}_i,\mathbf{u}_i\right)=\mathbf{f}_i\left(\mathbf{x}_i\right)+\mathbf{G}\mathbf{u}_i,
\end{equation}
where
\vspace{-0.25cm}
\[
\resizebox{0.99\hsize}{!}{%
$
{\color{black}\mathbf{f}_i\left(\mathbf{x}_i\right)=\begin{bmatrix}\dot{x}_i\\
    \dot{y}_i\\
    \dot{z}_i\\
    {f_i\over m_i}\left(\sin{\phi_i}\sin{\psi_i} + \cos{\phi_i}\cos{\psi_i}\sin{\theta_i}\right)\\
    {f_i\over m_i}\left(\cos{\phi_i}\sin{\psi_i}\sin{\theta_i}- \sin{\phi_i}\cos{\psi_i}\right)\\
    {f_i\over m_i}\cos{\phi_i}\cos{\theta_i}-g\\
    \dot{\phi}_i\\
    \dot{\theta}_i\\
    \dot{\psi}_i\\
    0\\
    0\\
    0\\
    {\color{black}\dot{f}}_i\\
    0\\
    % {\dot{p}_i\over m}\left(\sin{\phi_i}\sin{\psi_i} + \cos{\phi_i}\cos{\psi_i}\sin{\theta_i}\right)+{p_i\over m}\left[-\dot{\phi}_i\left(\cos{\psi_i} \sin{\phi_i} \sin{\theta_i} - \cos{\phi_i}\sin{\psi_i}\right)\right]\\
    % {\dot{p}_i\over m}\left(\cos{\phi_i}\sin{\psi_i}\sin{\theta_i}- \sin{\phi_i}\cos{\psi_i}\right)+{p_i\over m}\left[-\dot{\phi}_i\left(\cos{\phi_i}\cos{\psi_i} +\sin{\phi_i}\sin{\psi_i}\sin{\theta_i}\right)\right]\\
    % {\dot{p}_i\over m}\cos{\phi_i}\cos{\theta_i}+{p_i\over m}\left[-\dot{\phi}_i\cos{\theta_i}\sin{\phi_i}\right]\\
    \end{bmatrix}
    }
,
~
    {\mathbf{G}}    =\begin{bmatrix}
        \mathbf{g}_1&\cdots&\mathbf{g}_4
    \end{bmatrix}
    =
\begin{bmatrix}
\mathbf{0}_{9\times 1}&\mathbf{0}_{9\times 3}\\
\mathbf{0}_{3\times 1}&\mathbf{I}_3\\
0&\mathbf{0}_{1\times 3}\\
1&\mathbf{0}_{1\times 3}\\
\end{bmatrix}
.
$
}
\]
Note that $g=9.81 m/s^2$ is the gravity acceleration.
\vspace{-0.3cm}
\subsection{AVDC Trajectory Tracking}
The objective of trajectory tracking  is to choose  $\mathbf{u}_i$ such that $\mathbf{r}_{i,d}$ stably tracks the desired input $\mathbf{r}_{i,d}$. By applying the feedback linearization method developed in Refs. \cite{el2023quadcopter, rastgoftar2021safe}, we define state transformation
\vspace{-0.25cm}
\begin{equation}
    \mathbf{z}_i\left(\mathbf{x}_i\right)=\begin{bmatrix}    \mathbf{r}_i^T&\dot{\mathbf{r}}_i^T&\ddot{\mathbf{r}}_i^T&\dddot{\mathbf{r}}_i^T&\psi_i&\dot{\psi}_i
    \end{bmatrix}^T
\end{equation}
and let $\mathbf{z}_i$ be updated by the following dynamics:
\vspace{-0.25cm}
\begin{equation}\label{fedlindynamics}
    \dot{\mathbf{z}}=\mathbf{A}{\mathbf{z}}+\mathbf{B}\mathbf{w}_i,
\end{equation}
where $\mathbf{w}_i=\begin{bmatrix}
\ddddot{x}&\ddddot{y}&\ddddot{z}&\ddot{\psi}
\end{bmatrix}^T$ is the control input of external dynamics \eqref{fedlindynamics}, and 
\vspace{-0.25cm}
\[
    \mathbf{A}=\begin{bmatrix}
    \mathbf{0}_{9\times 3}&\mathbf{I}_9&\mathbf{0}_{9\times 1}&\mathbf{0}_{9\times 1}\\
    \mathbf{0}_{3\times 3}&\mathbf{0}_{3\times 9}&\mathbf{0}_{3\times 1}&\mathbf{0}_{3\times 1}\\
    \mathbf{0}_{1\times 3}&\mathbf{0}_{1\times 9}&0&1\\
    \mathbf{0}_{1\times 3}&\mathbf{0}_{1\times 9}&0&0\\
    \end{bmatrix}
    ~\mathrm{and}~
    % \in \mathbb{R}^{14\times 14}
    \mathbf{B}=\begin{bmatrix}
    \mathbf{0}_{9\times 3}&\mathbf{0}_{9\times 1}\\
    \mathbf{I}_{3}&\mathbf{0}_{3\times 1}\\
    \mathbf{0}_{1\times 3}&0\\
    \mathbf{0}_{1\times 3}&1\\
    \end{bmatrix}
    .
    % \in \mathbb{R}^{14\times 4}
    \]
Note that $\mathbf{w}_i$ is related to the control input of multi-copter UAS, denoted by $\mathbf{u}_i$, by
\vspace{-0.25cm}
\begin{equation}
    \mathbf{w}_i=\mathbf{M}_{i,1}\mathbf{u}_i+\mathbf{M}_{i,2},
\end{equation}
where
\begin{subequations}
\vspace{-0.2cm}
\begin{equation}
  \mathbf{M}_{i,1}=  \begin{bmatrix}
        L_{{\mathbf{g}}_{_{1}}}L_{{\mathbf{f}}_i}^{3}x& L_{{\mathbf{g}}_{_{2}}}L_{{\mathbf{f}}_i}^{3}x& L_{{\mathbf{g}}_{_{3}}}L_{{\mathbf{f}}_i}^{3}x& L_{{\mathbf{g}}_{_{4}}}L_{{\mathbf{f}}_i}^{3}x\\
        L_{{\mathbf{g}}_{_{1}}}L_{{\mathbf{f}}_i}^{3}y& L_{{\mathbf{g}}_{_{2}}}L_{{\mathbf{f}}_i}^{3}y& L_{{\mathbf{g}}_{_{3}}}L_{{\mathbf{f}}_i}^{3}y& L_{{\mathbf{g}}_{_{4}}}L_{{\mathbf{f}}_i}^{3}y\\
        L_{{\mathbf{g}}_{_{1}}}L_{{\mathbf{f}}_i}^{3}z& L_{{\mathbf{g}}_{_{2}}}L_{{\mathbf{f}}_i}^{3}z& L_{{\mathbf{g}}_{_{3}}}L_{{\mathbf{f}}_i}^{3}z& L_{{\mathbf{g}}_{_{4}}}L_{{\mathbf{f}}_i}^{3}z\\
         L_{{\mathbf{g}}_{_{1}}}L_{{\mathbf{f}}_i}\psi& L_{{\mathbf{g}}_{_{2}}}L_{{\mathbf{f}}_i}\psi& L_{{\mathbf{g}}_{_{3}}}L_{{\mathbf{f}}_i}\psi& L_{{\mathbf{g}}_{_{4}}}L_{{\mathbf{f}}_i}\psi\\
        \end{bmatrix}        
         \in \mathbb{R}^{4\times 4}
        .
\end{equation}
\vspace{-0.25cm}
\begin{equation}
  \mathbf{M}_{i,2}=  \begin{bmatrix}
        L_{{\mathbf{f}}_i}^{4}x&
       L_{{\mathbf{f}}_i}^{4}y&
        L_{{\mathbf{f}}_i}^{4}z&
         L_{{\mathbf{f}}_i}^2\psi
        \end{bmatrix}
        ^T\in \mathbb{R}^{4\times 1}
        .
\end{equation}
\end{subequations}
To achieve the control objective, we choose
\vspace{-0.25cm}
\begin{equation}
    \mathbf{w}_i=\mathbf{K}_i\left(\mathbf{z}_{i,d}-\mathbf{z}_i\right)
\end{equation}
such that $\mathbf{z}_i$ stably tracks 
\vspace{-0.25cm}
\[
\mathbf{z}_{i,d}=\begin{bmatrix}
    \mathbf{r}_{i,d}^T&\mathbf{0}_{1\times 9}&\psi_{d}&0
    \end{bmatrix}^T,
\]
where $\mathbf{r}_{i,d}\in \mathbb{R}^{3\times 1}$ given by Eq. \eqref{ridd} is the UAS  desired trajectory, and $\psi_d=0$ is the desired yaw angle of AVDC $i\in \mathcal{V}$. This control objective is achieved, if $\mathbf{A}-\mathbf{B}\mathbf{K}_i$ is Hurwitz, for every $i\in \mathcal{V}$. Then, the AVDC control is obtained by
\vspace{-0.25cm}
\begin{equation}
    \mathbf{u}_i=\mathbf{M}_{i,1}^{-1}\left(\mathbf{w}_i-\mathbf{M}_{i,2}\right). 
\end{equation}
Figure \ref{Blockdiagram} shows the block diagram of the proposed control applied by every AVDC $i\in \mathcal{V}$. Proofs for the stability and convergence of the proposed trajectory tracking control were provided in \cite{rastgoftar2021scalable, rastgoftar2021safe, el2023quadcopter}.
\vspace{-0.3cm}
\section{Simulation Results}\label{Simulation Results}
We consider an AVDC team identified by set $\mathcal{V}=\left\{1,\cdots,165\right\}$ with the reference configuration shown in Fig. \ref{InitialFormation}. We desire that the AVDC team provides computation resources over the separated rectangular and triangular domains shown in {\color{black}Fig. \ref{InitialFormation}}. Given the reference formation, the inter-agent communication is obtained and graphically shown in {\color{black}Fig. \ref{InitialFormation}}. This inter-AVDC communication links can be represented by the deep neural network shown in {\color{black}Fig. \ref{DNNStructure}}.

By knowing the DNN structure, the initial communication weight and final desired position of every AVDC are obtained by using the approach presented in Section \ref{AVDCs' Desired Positions}. The desired configuration of the AVDCs is shown by points in {\color{black}Fig. \ref{InitialFormation}}. By knowing the final desired positions of the agents, the final communication weights are obtained by \eqref{desiredrawwwwwwwwwwwwwwww} and the transient communication weigh is defined for every agent $i\in \mathcal{V}$ using Eq. \eqref{communicationweighttimevarying} over the time interval $\left[0,60s\right]$, where $\beta(t)$ is defined by a quintic polynomial and  plotted in {\color{black}Fig. \ref{beta}}.

To implement the proposed coverage approach, we assume every agent is a quadcopter UAS and apply the quadcopter model and trajectory tracking control developed in \cite{el2023quadcopter, rastgoftar2021safe} to operate DNN neurons. Figure \ref{IntemediateConfs} plots the intermediate configuration of the AVDC team at sample times $t=15s$, $t=35s$, and $t=80s$ where the actual path of every AVDC $i\in \mathcal{V}$, from $\mathbf{a}_i$ to $\mathbf{p}_i$, is shown by dashed red.

\vspace{-0.3cm}

\section{Conclusion}\label{Conclusion}
Our research approach will utilize mobile application-aware management system (MAMS) that is aware of both the application characteristics and the underlying UAS infrastructures to break the barriers between applications, middleware/operating system, and UAS hardware layers. The vertical integration of these layers will enable us to build a dynamically customizable Aerial Virtual Data Center (AVDC) that will be optimized to meet the mobile application requirements such as QoS, availability, resilience, and energy. Our approach to develop the proposed Aerial-based Mobile Composable Data center (AMCDC) is shown in Figure 1 that benefits from virtualization techniques in a variety of ways. Virtual Machines (VMs) have been widely used to provide a layer that is well-positioned in the hardware/software stack of computer systems. This provides fine-grain resource monitoring and control capabilities. In the AMCDC approach, we scale the virtualization concept from being at the VM level to an Aerial Virtual Data Center (AVDC) level. For a given mobile application class, the AMCDC middleware can build a AVDC that can meet the application Service Level Objectives (SLO) such as security, performance, energy consumption, resilience, or availability objectives. The selected AVDC will be then mapped into a set of available UAS resources that will provide the required computation, storage, communications requirements. The composable UAS building blocks include heterogeneous drones that provide general purpose graphics processing units (GPUs), general purpose processors, storage capabilities and communications services.

\vspace{-0.3cm}

\bibliographystyle{IEEEtran}
\bibliography{reference}

\end{document}